\documentclass[runningheads]{llncs}

\usepackage{xcolor}
\usepackage[hidelinks,colorlinks = true,
           linkcolor = blue,
            urlcolor  = blue,
            citecolor = blue,
            anchorcolor = blue,
            bookmarks=false]{hyperref}
\usepackage{graphicx,subcaption}
\usepackage{enumitem}
\usepackage{amsmath}
\usepackage[noend]{algorithmic}
\usepackage{algorithm}
\usepackage{tikz}
\usepackage[labelfont=bf]{caption}
\usepackage{multicol}
\usepackage{multirow}
\usepackage{array}
\usepackage{booktabs}
\newcolumntype{L}[1]{>{\raggedright\let\newline\\\arraybackslash\hspace{0pt}}m{#1}}
\newcolumntype{C}[1]{>{\centering\let\newline\\\arraybackslash\hspace{0pt}}m{#1}}
\newcolumntype{R}[1]{>{\raggedleft\let\newline\\\arraybackslash\hspace{0pt}}m{#1}}

\usepackage{makecell}
\usepackage{pgfplots,pgfplotstable}
\usepackage{graphics}
\usepackage{tikz}
\usepackage{circuitikz}
\usetikzlibrary{bayesnet}
\tikzstyle{graphnode}=[circle,draw, fill=white, minimum size=2em]
\tikzstyle{decision} = [diamond, draw, text badly centered, inner sep=3pt]
\usetikzlibrary{matrix}
\usetikzlibrary{chains}
\usetikzlibrary{arrows.meta,calc,chains,shapes.geometric}
\usetikzlibrary{backgrounds,fit,intersections,patterns,pgfplots.fillbetween}
\usetikzlibrary{bayesnet}
\definecolor{myblue}{rgb}{0.05, 0.15, 0.6}
\definecolor{purple}{rgb}{0.54, 0.17, 0.89}
\definecolor{darkgreen}{rgb}{0.0, 0.5, 0.0}

\definecolor{darkbrown}{rgb}{0.4, 0.26, 0.13}

\definecolor{amber}{rgb}{1.0, 0.75, 0.0}
\definecolor{bblue}{rgb}{0.13, 0.67, 0.8}

\usepackage{longtable}
\usepackage{array}
\newcolumntype{L}[1]{>{\raggedright\let\newline\\\arraybackslash\hspace{0pt}}m{#1}}
\newcolumntype{C}[1]{>{\centering\let\newline\\\arraybackslash\hspace{0pt}}m{#1}}
\newcolumntype{R}[1]{>{\raggedleft\let\newline\\\arraybackslash\hspace{0pt}}m{#1}}
\usepackage{multicol}

\newcommand{\dmanthan}{{Manthan3}}

\newcommand{\hqs}{HQS2}
\newcommand{\pedant}{Pedant}

\newcommand{\manthan}{\ensuremath{\mathsf{Manthan}}}

\newcommand{\unsat}{{UNSAT}}
\newcommand{\unsatcore}{{UnsatCore}}
\newcommand{\sat}{{SAT}}
\newcommand{\maxsat}{{MaxSAT}}

\newcommand{\totalorder}{\ensuremath{\mathsf{Order}}}

\newcommand{\candidateskf}{\ensuremath{\mathsf{CandidateHkF}}}

\newcommand{\findorder}{\ensuremath{\mathsf{FindOrder}}}
\newcommand{\checksat}{\ensuremath{\mathsf{CheckSat}}}
\newcommand{\refine}{\ensuremath{\mathsf{RepairHkF}}}
\newcommand{\substitute}{\ensuremath{\mathsf{Substitute}}}
\newcommand{\getsamples}{\ensuremath{\mathsf{GetSamples}}}

\newcommand{\createdt}{\ensuremath{\mathsf{CreateDecisionTree}}}

\newcommand{\Path}{\ensuremath{\mathsf{Path}}}

\newcommand{\Label}{\ensuremath{\mathsf{Label}}}

\newcommand{\findcore}{\ensuremath{\mathsf{FindCore}}}
\newcommand{\maxsatlist}{\ensuremath{\mathsf{FindCandi}}}

\usepackage[draft]{todonotes}   %

\begin{document}
\title{Synthesis with Explicit Dependencies\thanks{ To be appeared at DATE 2023. {\dmanthan} is available open-sourced at \url{https://github.com/meelgroup/manthan}.}}

\institute{}
\author{Priyanka Golia\inst{1,2} \and  Subhajit Roy\inst{1} \and Kuldeep S. Meel\inst{2}}
\authorrunning{Golia, Roy, and Meel}
\institute{Indian Institute of Technology Kanpur, India  \and  National University of Singapore, Singapore}

\maketitle             

	\begin{abstract}

Quantified Boolean Formulas (QBF) extend propositional logic with quantification $\forall, \exists$. In QBF, an existentially quantified variable is allowed to depend on all universally quantified variables in its scope. Dependency Quantified Boolean Formulas (DQBF) restrict the dependencies of existentially quantified variables. In DQBF, existentially quantified variables have explicit dependencies on a subset of universally quantified variables, called Henkin dependencies. Given a Boolean specification between the set of inputs and outputs, the problem of Henkin synthesis is to synthesize each output variable as a function of its Henkin dependencies such that the specification is met. Henkin synthesis has wide-ranging applications, including verification of partial circuits, controller synthesis, and circuit realizability.

This work proposes a data-driven approach for Henkin synthesis called {\dmanthan}. On an extensive evaluation of over 563 instances arising from past DQBF solving competitions, we demonstrate that {\dmanthan} is competitive with state-of-the-art tools. Furthermore, {\dmanthan} could synthesize Henkin functions for 26 benchmarks for which none of the state-of-the-art techniques could synthesize.
\end{abstract}

	\section{Introduction}\label{sec:intro}
Quantified Boolean Formulas (QBF) equip the propositional logic with universal $(\forall)$ and existential quantifiers $(\exists)$ for propositional variables. In QBF, an existentially quantified variable is allowed to depend on all universally quantified variables within its scope. On the other hand, Henkin quantifiers, often called Branching quantifiers, generalize the standard quantification and allow explicit declarations of dependencies~\cite{KM95}. Propositional logic is equipped with Henkin quantifiers, resulting in the so-called Dependency Quantified Boolean Formulas (DQBFs). In DQBF, an existentially quantified variable is allowed to depend on a pre-defined subset of universally quantified variables, called Henkin dependencies. For example, $\phi: \forall x_1,x_2 \exists^{x_1} y_1. \varphi(x_1,x_2,y_1)$ is a DQBF formula, where $\varphi$ is some quantifier-free Boolean formula and existentially quantified variable $y_1$ is only allowed to depend on $x_1$, which is the Henkin dependency corresponding to $y_1$. The dependency specification quantification is called Henkin quantifier~\cite{KM95}.

These explicit dependencies provide more succinct descriptive power to DQBF than QBF. However, DQBF is shown to be in the complexity class of NEXPTIME-complete~\cite{PRA01}, whereas QBF is \emph{only}  PSPACE-complete~\cite{M79}. The payoffs associated with an increase in the computational complexity are the wide-ranging applications of DQBF, such as engineering change of order~\cite{JKL20}, topologically constrained synthesis~\cite{BKJ14}, equivalence checking of partial functions~\cite{GRSWSB13}, finding strategies for incomplete games~\cite{PRA01}, controller synthesis~\cite{BKS14}, circuit realizability~\cite{BKJ14}, and synthesis of fragments of linear-time temporal logic~\cite{CHOP13}.

The DQBF satisfiability is a decision problem that looks for an answer to the question: \emph{Does there exist a function corresponding to each existentially quantified variable, in terms of its Henkin dependencies, such that the formula substituted with the function in places of existentially quantified variables is a tautology?} Owing to wide variety of applications that can be represented as DQBF, recent years have seen a surge of interest in DQBF solving~\cite{GRSWSB13,FKB14,RSS21,TR19,WWSB16}. 
In many cases, a mere True/False answer is not sufficient as one is often interested in determining the definitions corresponding to those functions. For instance, in the context of engineering change of order (ECO), in addition to just knowing whether the given circuit could be rectified to meet the \emph{golden} specification, one would also be interested in deriving corresponding patch functions~\cite{JKL20}. Owing to the naming of dependencies, we call such patch functions to be Henkin functions.

Recent years have witnessed an increased interest in the problem of Henkin function synthesis. The state-of-the-art techniques, {\hqs}~\cite{WWSB16} and {\pedant}~\cite{RSS21} can synthesize Henkin functions for True DQBF in addition to DQBF solving. {\hqs}~\cite{WRMB17} applies a sequence of transformations to eliminate quantifiers in DQBF instances to synthesize Henkin functions for True instances, whereas {\pedant}~\cite{RSS21} uses interpolation-based definition extraction and various SAT oracle calls to synthesize Henkin functions. Despite the significant progress over the years, many real-world instances are beyond the reach of Henkin function synthesis engines.  

In this work, we take a step to push the envelope of Henkin synthesis. To this end, we propose a novel framework for Henkin function synthesis, called {\dmanthan}. {\dmanthan} takes an orthogonal approach to the existing techniques by combining advances in machine learning with automated reasoning. In particular, {\dmanthan} uses constrained sampling to generate the data, which is later fed to a machine-learning algorithm to learn the candidate functions in accordance with the Henkin dependencies for each existentially quantified variable. Then, {\dmanthan} employs a SAT solver to check the correctness of the synthesized candidates. If the candidate verification checks fail, {\dmanthan} does a counterexample-driven candidate repair. Furthermore, {\dmanthan} utilizes a {\maxsat} solver-based method to find the candidates that need to undergo repair and uses a proof-guided strategy to construct a \emph{good} repair.

To demonstrate the practical efficiency of {\dmanthan}, we perform an extensive comparison with the prior state-of-the-art techniques, {\hqs} and {\pedant}, over a benchmark suite of 563 instances. Our empirical evaluation demonstrates that {\dmanthan} shows competitive performance and significantly contributes to the portfolio of Henkin synthesizers. {\dmanthan} achieves the shortest synthesizing time on 42 of the 204 benchmarks solved by at least one tool. Furthermore, {\dmanthan} is able to synthesize Henkin functions for 26 instances that none of the state-of-the-art function synthesis engines could synthesize.

	\section{Preliminaries}\label{sec:prelim}
We use a lower case letter to represent a propositional variable and an upper case letter to represent a set of variables. A literal is either a variable or its negation, and a clause is considered as a disjunction of literals. 
A formula $\varphi$ represented as conjunction of clauses is considered in Conjunctive Normal Form (CNF). \emph{Vars}($\varphi$) represents the set of variables appearing in $\varphi$. A satisfying assignment$(\sigma)$ of the formula $\varphi$ maps \emph{Vars}($\varphi)$ to $\{0,1\}$ such that $\varphi$ evaluates to True under $\sigma$. We use $\sigma \models \varphi$ to represent $\sigma$ as a satisfying assignment of $\varphi$. For a set of variables $V$, we used $\sigma[V]$ to denote the restriction of $\sigma$ to $V$. If $\varphi$ evaluates to True for all possible valuation of \emph{Vars}($\varphi$), $\varphi$ is considered as tautology.

A uniform sampler samples the required number of satisfying assignments uniformly at random from the solution space of the formula. We use {\unsatcore} to represents an $\emph{unsatisfiable core}$, which is a subset of clauses of $\varphi$ for which there does not exists a satisfying assignment. For a CNF formula in which a set of clauses is considered as hard constraints and remaining clauses as soft constraints, a {\maxsat} solver tries to find a satisfying assignment that satisfies all hard constraints and maximizes the number of satisfied soft constraints.

A formula $\phi$ is DQBF if it can be represented as $\phi:  \forall x_1 \ldots x_n$  $\exists^{H_1} y_1 \ldots \exists^{H_m} y_m$ $\varphi(X,Y)$ where $X = \{x_1,\ldots,x_n\}$, $Y = \{y_1,\ldots,y_m\}$ and $H_i \subseteq  X$ represents the dependency set of $y_i$, that is, variable $y_i$ can only depend on $H_i$. Each $H_i$ is called Henkin dependency and each quantifier $\exists^{H_i}$ is called \emph{Henkin} quantifier~\cite{H61}. 

A DQBF $\phi$ is considered to be True, if there exists a function $f_i: \{0,1\}^{|H_i|} \mapsto \{0,1\}$ for each existentially quantified variable $y_i$, such that $\varphi(X,f_1(H_1),$ $\ldots,f_m(H_m))$, obtained by substitution of each $y_i$ by its corresponding function $f_i$, is a tautology. Given a DQBF $\phi$, the problem of DQBF satisfiability, is to determine whether a given DQBF is True or False. \\

{\bfseries Problem Statement:} Given a True DQBF $\forall x_1 \ldots x_n$ $\exists^{H_1} y_1 \ldots \exists^{H_m} y_m$ $\varphi(x_1, \ldots$ $, x_n,y_1, \ldots, y_m) $ where $ x_1,\ldots,x_n$ $\in X$, $y_1,\ldots,y_m\in Y$, $H_i \subseteq X$, the problem of {\bfseries Henkin Synthesis}  is to synthesize a function vector $ \boldsymbol{f}: \langle f_1,\dots, f_m \rangle$ such that 
$\varphi(X,f_1(H_1),$ $\ldots,f_m(H_m))$ is a tautology.  

$\boldsymbol{f}$ is called Henkin function vector and each $f_i$ is a Henkin function.  We used $\forall X \exists^{H_1} y_1 \ldots  \exists^{H_m} y_m$ $ \varphi(x_1,\ldots,x_n,$ $y_1,\ldots,y_m)$ and $\forall X \exists^{H} Y \varphi(X,Y)$ interchangeably.

Henkin synthesis generalizes Skolem synthesis in which $H_1 = \ldots = H_m = X$. In such a case, one omits the usage of $H_i$ and simply represents $\phi$ as $\forall X \exists Y \varphi(X,Y)$. In such a case, $\boldsymbol{f}$ is called Skolem function vector, such that $\forall X (\exists Y \varphi(X,Y) \leftrightarrow \varphi(X, \boldsymbol{f}))$.

\section{Related Work}~\label{sec:relatedwork}

There has been a lot of work in the area of QBF solving~\cite{J18,J18a,LB10,LE17,R19,RT15,RTRS18}, and the recent years have seen significant interest in DQBF solving as well ~\cite{FKB14,GRSWSB13,GWRSSB15,J20,RSS21,TR19,WWSB16}. The first DPLL-based approach to solve DQBF satisfiability was proposed by Frohlich et al.~\cite{FKB12}. In a similar direction, recently,  Tentrup and Rabe~\cite{TR19} introduced the idea of using clausal abstraction for {DQBF} solving. Gitina et al.~\cite{GRSWSB13} proposed the idea of solving a DQBF instance using a basic variable elimination strategy which is about transforming a DQBF instance to a QBF by eliminating a set of variables that causes non-linear dependencies. The strategy is further improved by several optimization in~\cite{GWRSSB15}. A similar approach was proposed by Frohlich et al.~\cite{FKB14} in which a DQBF instance is transformed into an SAT instance by a local universal expansion on each clause. 

Wimmer et al.~\cite{WWSB16} proposed a method for obtaining Henkin functions from DQBF solvers that are based on variable elimination-based DQBF solving techniques~\cite{GWRSSB15,FKB14}. Elimination-based DQBF solvers execute a sequence of transformation of eliminating quantifiers on a DQBF instance to obtain an SAT instance; in this process, they obtain a sequence of equisatisfiable formulas $\varphi_1, \varphi_2, \ldots, \varphi_k$, where $\varphi_i$ formula is a result of a transformation of $\varphi_{i-1}$. They showed that for a True DQBF instances, Henkin function for $\varphi_{i-1}$ can be obtained from $\varphi_{i}$. 

Recently, Reichl, Slivovsky, and Szeider~\cite{RSS21} proposed a different approach for DQBF solving using interpolation-based definition extraction. They introduce \emph{arbiter variables} that represents the value of an existential variable for all the assignments of its dependency sets for which the variable is not uniquely defined. And, then it extracts the definition for existential variables in terms of their dependency sets and those arbiter variables. The proposed method is certifying by design and returns Henkin functions. 

The earliest work in the context of DQBF focused on the special case of 2-QBF for both decision and functional synthesis problems. The problem of functional synthesis in the context of 2-QBF is known as Skolem synthesis. The earliest works on Skolem synthesis were limited to the case when the given 2-QBF formula was True~\cite{BJ11,HSB14,J09}; the subsequent work expanded to handle general 2-QBFs, and consequently, a multitude of wide-ranging techniques have been proposed, such as CEGAR-based methods~\cite{AACKRS19,ACGKS18,FTV16}, factorization-based techniques~\cite{ACJS17,JSCTA15}, incremental determinization~\cite{R19}.

Recently, a new data-driven approach for Skolem synthesis was implemented in {\manthan}~\cite{GRM20,GSRM21}. {\manthan} uses advances in constraint sampling, machine learning, and formal methods for efficient Skolem synthesis. {\manthan} uniformly samples from the solutions of $\varphi$. Using these samples as data, it learns a candidate vector for Skolem functions as a decision tree classifier for each existentially quantified variable. {\manthan}, then, verifies whether the learned candidate vector is a Skolem vector. If the candidate vector passes the verification check, {\manthan} returns the  Skolem vector. Otherwise, {\manthan} uses the counterexample to perform an {\unsatcore} guided repair and relies on a {\maxsat} solver to minimize the number of candidates to repair in each counterexample.

\section{Overview}
\label{sec:depmanthan}

This section provides a high-level overview of {\dmanthan} framework. While {\dmanthan} shares high-level similarity with {\manthan}, the recently proposed Skolem function synthesis engine ~\cite{GRM20}\cite{GSRM21}, in its usage of machine learning techniques and SAT/MaxSAT solvers,  the two techniques differ crucially due to the requirements imposed by Henkin dependencies. It is worth remarking that handling Henkin dependencies is not trivial, perhaps best highlighted by the fact that 2-QBF is $\Sigma_2^P$-complete while DQBF is NEXPTIME-complete~\cite{PRA01}.

As shown in Figure~\ref{fig:maindia}, {\dmanthan} first uses advances in constrained sampling to generate the data, then use the data to learn a candidate vector $\boldsymbol{f}$ using a machine learning-based approach. Then, {\dmanthan} attempts to verify if the candidate vector $\boldsymbol{f}$ is a Henkin function vector. If the candidates pass the formal verification check, {\dmanthan} returns the candidates as a valid Henkin vector. Otherwise, the candidate vector is repaired to satisfy the counterexample, and the verification check is repeated. Note that {\dmanthan} needs to take care of restrictions imposed by Henkin dependencies while learning and repairing the candidates.

We now present high-level overview of the different components of {\dmanthan}:

\begin{figure*}
    \centering

\begin{tikzpicture}[%
   scale = 0.5, node distance=5mm
    ,>=latex'
    ,block/.style = {%
        ,draw
        ,minimum height=10mm,minimum width=28mm
        ,align=center
        }
    ,start chain = going below,
    arrow/.style = {thick,-{Stealth[length=5pt,width=4pt]}},
    decoration={brace,amplitude=4pt},scale=1]
    \tikzstyle{bigbox} = [draw=black, thick, fill=blue!10, rectangle,inner xsep=0.2cm, inner ysep=0.25cm]
    
    \node [block,fill=bblue!60] (dg) {Data Generation};
    \node [left = of dg,xshift = -0.2cm] (oi) {};
    \node [left =of dg, xshift = 1cm, align = left,text width=2cm] (i) {Input $\varphi({X},{Y})$};
	\node [block, fill=red!40,  right = of dg] (lsf) {Learn Candidate\\ Functions};
  {
    \node [draw, diamond, aspect=3, fill=amber!50, below = of lsf] (v) {Verify};
    \node [block, fill=amber!50, right = of v,minimum height=10mm,minimum width=20mm] (r) {Repair};
    \node [right =of r,yshift = -0.8cm,xshift=0.20cm] (o) {Output $\boldsymbol{f}$};
}

    \node [below =of v,yshift = 0.4cm] (t) {};
  
    \draw[->] (oi) to (dg);
  {  \draw[->] (dg) to (lsf);}
    {  \draw[->] (lsf) to (v);
    \path[->,draw]
    (v) edge node [above] {{No}} (r)
    (r) edge[below, bend left] (v);
    \draw[arrow,thin] (v.south) -- node[left, near start] {Yes} ++(0,-0.75) |-(16,-4.6) (o);
}

	\begin{pgfonlayer}{background}
	\node[bigbox] [fit = (dg) (t) (r)] (b){};
	\end{pgfonlayer}

\end{tikzpicture}
\caption{Overview of {\dmanthan}}~\label{fig:maindia}
\end{figure*}

{\bfseries Data Generation:} As the first step, {\dmanthan} uses constrained samplers~\cite{GSCM21,GSRM19} to sample the satisfying assignments of specification $\varphi$ uniformly at random from the solution space of specification. The sampled satisfying assignments are considered data to feed the learning algorithms to learn candidate functions in the next stage.

 {\bfseries Candidate Learning:} {\dmanthan} learns a binary decision tree classifier for each existentially quantified variable $y_i$ to learn the candidate function $f_i$ corresponding to it. The valuations of $y_i$ in the generated samples are considered labels, and the valuations of corresponding Henkin dependencies $H_i$ are considered the feature set to learn a decision tree. A Henkin function $f_i$ corresponding to $y_i$ is computed as a disjunction of labels along all the paths from the root node to leaf nodes with label $\boldsymbol{1}$ in the learned decision tree. 

Due to the Henkin dependencies, the feature set for $y_i$ must be restricted only to $H_i$. However, in order to learn a good decision tree, we can include all the $y_j$ in the set of features for which $H_j \subset H_i$. The function $f_j$ can be simply expanded within $f_i$ so that $f_i$ is only expressed in terms of $H_i$. For the cases when $H_j = H_i$, such use of the $Y$ variables is allowed as long it does not cause the cyclic dependencies; that is, if $y_j$ appears in the learned candidate $f_i$, then $y_i$ is not allowed as a feature to learn candidate $f_j$. If $y_j$ appears in $f_i$, then we say $y_i$ depends on $y_j$, denoted as $y_i \prec_{d} y_j$. {\dmanthan} discovers requisite variable ordering constraints among such $Y$ variables on the fly as the candidate functions are learned.

A function vector $\boldsymbol{f}$ in which $y_j$ appears in $f_i$ is a valid vector if $y_i$ does not appear in $f_j$. If $\boldsymbol{f}$ is a valid function vector, there exists a partial order $\prec_{d}$ over $\{y_1,\ldots,y_{m}\}$. Once, we have a candidate vector, {\dmanthan} obtains a valid linear extension total order, say denoted as {\totalorder}, from the partial dependencies learned in \textit{candidate learning} over $Y$ variables.

 {\bfseries Verification:} The learned candidate vector may not always be a valid Henkin vector. Therefore, the candidate functions must be verified. $\boldsymbol{f}$ is a Henkin function vector only if $\varphi(X,f_1(H_1),$ $\ldots,f_m(H_m))$ is a tautology. {\dmanthan} first, make a SAT oracle query on the formula \label{eqn:errorformula} $E(X,Y') = \lnot \varphi(X,Y') \land (Y' \leftrightarrow \boldsymbol{f})$

If formula $E(X,Y')$ is {\unsat}, {\dmanthan}  returns the function vector $\boldsymbol{f}$ as a Henkin function vector. If formula $E(X,Y')$ is SAT and $\delta$ is a satisfying assignment of $E(X,Y')$, {\dmanthan} needs to find out whether $\varphi(X,Y)$ has a propositional model extending assignment of $X$. Therefore, {\dmanthan} performs another satisfiability check on formula $\varphi(X,Y) \land (X \leftrightarrow \delta[X])$. If satisfiability checks return UNSAT, the corresponding $DQBF$ formula is False, and there does not exist a Henkin function vector; therefore, {\dmanthan} terminates. Furthermore, if $\varphi(X,Y) \land (X \leftrightarrow \delta[X])$ is SAT, and $\pi$ is a satisfying assignment and we need to repair the candidate function vector. Note that $\pi[X]$ is same as $\delta[X]$, and $\pi[Y]$ is a possible extending assignment of $X$, and $\delta[Y']$ presents the output of candidate function vector with $\delta[X]$. Now, we have a counterexample $\sigma$ as $\pi[X] + \pi[Y] + \delta[Y']$.

{\bfseries Candidate Repair:} We apply a counterexample driven repair approach for candidate functions. As {\dmanthan} attempts to fix the counterexample $\sigma$, it first needs to find which candidates to repair out of $f_1$ to  $f_m$ candidates. {\dmanthan} takes help of {\maxsat} solver to find out the repair candidates, and makes a {\maxsat} query with $\varphi(X,Y) \land (X \leftrightarrow \sigma[X])$ as hard constraints and $(Y \leftrightarrow \sigma[Y'])$ as soft constraints. It selects a function $f_i$ for repair if the corresponding soft constraint $y_i \leftrightarrow \sigma[y_i']$ is falsified in the solution returned by the {\maxsat} solver. Once, we have candidates to repair, {\dmanthan} employs unsatisfiability cores obtained from the infeasibility proofs capturing the reason for candidates to not meet the specification to construct a repair.

Let us now assume that {\dmanthan} selects $f_i$ corresponding to variable $y_i$ as a potential candidate.  {\dmanthan} constructs another formula $G_i(X,Y)$ (Formula \ref{eqn:Gformuladqbf}) to find the repair:
\begin{multline}
\label{eqn:Gformuladqbf}
G_{i}(X,Y):\varphi(X,Y) \land (H_i \cup \hat{Y} \leftrightarrow \sigma[H_i \cup \hat{Y}])  \land (y_i \leftrightarrow \sigma[y'_{i}]) \\
\text{ where } \hat{Y} \subseteq Y \text{ such that } \forall y_j \in
 \hat{Y}:  H_j \subseteq H_i \\ \text{ and } 
\{ \totalorder[index(y_j)] > \totalorder[index(y_i)]\} 
\end{multline}

Informally, in order to determine whether $f_i$ needs to be repaired, we conjunct the specification $\varphi(X,Y)$ with the conjunction of unit clauses that set the valuation of $y_i$ to the current output of $f_i$ and the valuation of all the dependencies as per the counter-example. We describe the intuition behind construction of $G_{i}(X,Y)$. The formula $G_i(X,Y)$ is constructed to answer the following question: {\em Whether is it possible for $y_i$ to be set to the output of $f_i$ given the valuation of its Henkin dependencies?}. 

 The answer to the above question depends on whether $G_i(X,Y)$ is {\unsat} or {\sat}.  
$G_i(X,Y)$ being {\unsat} indicates that it is not possible for $y_i$ to be set to the output of $f_i$ and the {\unsatcore} of $G_i(X,Y)$ captures the reason. Accordingly,  {\dmanthan} uses the {\unsatcore} of $G_i(X,Y)$ to repair the candidate function $f_i$. In particular, {\dmanthan} uses all 
the variables corresponding to unit clauses in {\unsatcore} of $G_i(X,Y)$ to construct a repair formula $\beta$, and depending on the valuation of $y'_i$ in the counter example $\sigma$, $\beta$ is used to strengthen or weaken the candidate $f_i$ to satisfy the counterexample.

On the other hand, if $G_i(X,Y)$ is {\sat}, {\dmanthan} attempts to find alternative candidate functions to repair. $G_i(X, Y)$ being {\sat} indicates that with the current valuation to Henkin dependencies, $y_i$ could take a value as per the output of candidate $f_i$; however, to fix the counterexample $\sigma$, we need to repair another candidate function. To this end, let  $\rho$ be a satisfying assignment of $G_i(X, Y)$, then all $y_j$ variables for which $\rho[y_j]$ is not the same as $\sigma[y'_j]$ are added to the queue of potential candidates to repair.

The repair loop continues until either $E(X,Y')$ is {\unsat} or $\varphi(X,Y) \land (X \leftrightarrow \delta[X])$ is {\unsat}, where $\delta$ is a satisfying assignment of $E(X,Y')$ . If  $E(X,Y')$ is {\unsat}, we have a Henkin function vector $\boldsymbol{f}$, and if $\varphi(X,Y) \land (X \leftrightarrow \delta[X])$ is {\unsat}, then the given DQBF instance is False and there does not exist a Henkin function vector.

	\section{Algorithmic Details}~\label{sec:algo}

{\dmanthan} (Algorithm~\ref{algo:main_algo}) takes a DQBF instance $\forall X \exists^{H_1} y_1$ $\ldots \exists^{H_m} y_m \varphi(X,Y)$ as input and outputs a Henkin function vector $\boldsymbol{f}:=\langle f_1,\ldots,f_m \rangle$. 
\begin{algorithm}
	\footnotesize
	\caption{{\dmanthan}($\forall X \exists^{H} Y. \varphi(X,Y)$)}
	\label{algo:main_algo}
\begin{algorithmic}[1]

\STATE $\Sigma \gets$ {\getsamples}($\varphi(X,Y)$) \label{mainalgo:line:getsample} 

\STATE $D \gets \{d_1 =\emptyset \ldots, d_{|Y|} =\emptyset\}$\label{mainalgo:line:d}

\FOR {{$\langle H_i, H_j\rangle$}\label{mainalgo:line:d_i}}
{
	\IF {{$H_j \subset H_i$}\label{mainalgo:line:d_i-1}}
		{
			\STATE {$d_j \gets d_j \cup y_i$\label{mainalgo:line:d_i-2}}
		}
	\ENDIF	
}
\ENDFOR

\FOR{$y_i \in {Y}$ \label{mainalgo:line:candidate-learn}}
{
	
    \STATE $f_i,D\gets${\candidateskf}($\Sigma,\varphi(X,Y),y_i,D$)\label{mainalgo:line:phase1end}
	
}
\ENDFOR

\STATE $\totalorder \gets$ {\findorder}($D$) \label{mainalgo:line:findorder} 

\REPEAT 
{
	\STATE $E(X,Y') \gets  \lnot \varphi(X,Y') \land (Y'\leftrightarrow \boldsymbol{f}) $
	\label{mainalgo:line:refinestart}
	
	\STATE $ret, \delta \gets$ {\checksat}($E(X, Y')$) \label{mainalgo:line:checksat} 

	\IF{ret = SAT}
	
			\STATE $res, \pi \gets$ {\checksat}($\varphi(X,Y) \land (X \leftrightarrow \delta[X])$)  \label{mainalgo:line:checksat2}
			
			\IF{res = UNSAT}
			\RETURN $\forall X \exists^{H} Y. \varphi(X,Y)$ is False. \label{mainalgo:line:checksat2-unsat}
			
			\ENDIF
			
			\STATE $\sigma \gets \pi[X] + \pi[Y] + \delta[Y']$ \hfill \COMMENT{$\sigma$ is a counterexample} \label{mainalgo:line:checksat2-cex}

			\STATE $\boldsymbol{f} \gets {\refine}(\varphi(X,Y),\boldsymbol{f},\sigma,{\totalorder})$  \label{mainalgo:line:refine}
	\ENDIF
		
}\UNTIL{ret = UNSAT}
\STATE $\boldsymbol{f} \gets$ {\substitute}($\varphi(X,Y),\boldsymbol{f},{\totalorder}$)\label{mainalgo:line:substitute}
\RETURN $\boldsymbol{f}$
\end{algorithmic}
\end{algorithm}

Algorithm~\ref{algo:main_algo} assumes access to the following subroutines:

\begin{enumerate}
	\item {\getsamples}: It takes a specification as input and calls an oracle to produce samples $\Sigma$ of specifications. Each sample in $\Sigma$ is a satisfying assignment of specifications.
	\item {\candidateskf}: This subroutine generates the candidate function corresponding to an existential variable. It takes a specification $\varphi$, generated samples $\Sigma$, existential variable $y_i$ corresponding to which we want to learn a candidate function and a vector $D$ that keeps track of dependencies among $Y$ variables as input. {\candidateskf} returns a candidate function $f_i$  corresponding to $y_i$, and updates the dependencies in $D$ for $y_i$. We discussed {\candidateskf} routine in detail in Algorithm~\ref{decision_tree_skf_algo}.
	\item {\findorder}: It takes a set $D$ collection of $d_i$, where each $d_i$ is the list of $Y$ variables, which can depend on $y_i$. {\findorder} obtains a valid linear extension, {\totalorder}, from the partial dependencies  in $D$. 
	
	\item {\checksat}: It takes a specification as input and makes a SAT oracle call to do a satisfiability check on the specification. It returns the outcome of satisfiability check as {\sat} or {\unsat}. In the case of {\sat}, it also returns a satisfiable assignment of the specification.
	
	\item {\refine}: This subroutine repairs the current candidate function vector to fix the counterexample. It takes the specification, candidate function vector, a counterexample, and {\totalorder}, a linear extension of dependencies among $Y$ variables as input, and returns a repaired candidate function vector. Algorithm~\ref{refine_algo} discusses {\refine} subroutine in detail.
\end{enumerate}

Algorithm~\ref{algo:main_algo} starts with generating samples $\Sigma$ by calling {\getsamples} subroutine at line~\ref{mainalgo:line:getsample}. Next, Algorithm~\ref{algo:main_algo} initializes the set $D$ (line~\ref{mainalgo:line:d}), which is a collection of $d_i$, where $d_i$ represents the set of $Y$ variables that depends on $y_i$. Lines~\ref{mainalgo:line:d_i}-\ref{mainalgo:line:d_i-2} introduce variable ordering constraints based on the subset relations in each $\langle H_i, H_j \rangle$ pair, that is, if $H_j \subset H_i$, then $y_i$ can depend on $y_j$. Line~\ref{mainalgo:line:phase1end} calls the subroutine {\candidateskf} for every $y_i$ variable to learn the candidate function $f_i$. Next, at line~\ref{mainalgo:line:findorder}, {\dmanthan} calls {\findorder} to compute {\totalorder}, a topological ordering among the $Y$ variables that satisfy all the ordering constraints in $D$. 

In line~\ref{mainalgo:line:checksat}, {\checksat} checks the satisfiability of the formula $E(X,Y')$ described at line~\ref{mainalgo:line:refinestart}. If $E(X,Y')$ is {\sat}, then {\dmanthan} at line~\ref{mainalgo:line:checksat2} performs another satisfiability check to ensure that propositional model to $X$ can be extended to $Y$. If {\checksat} at  line~\ref{mainalgo:line:checksat2} is UNSAT, then Algorithm~\ref{algo:main_algo} terminates at line~\ref{mainalgo:line:checksat2-unsat} as there does not exists a Henkin function vector, otherwise {\dmanthan} has a counterexample $\sigma$ to fix.  The candidate vector $\boldsymbol{f}$ goes into a repair iteration (line~\ref{mainalgo:line:refine}) based on the counterexample $\sigma$, that is, the subroutine {\refine} repairs the current function vector $\boldsymbol{f}$ such that $\sigma$ now gets fixed.  {\dmanthan} returns a function vector $\boldsymbol{f}$ only if $E(X,Y')$ is {\unsat}.

\begin{algorithm}

\caption{{\candidateskf}$(\Sigma,\varphi(X,Y),y_i,D)$}
\label{decision_tree_skf_algo}

\begin{algorithmic}[1]

\STATE {$featset \gets H_i$} \label{mainalgo:line:featset-hi}
\FOR { {$y_j \in Y$} \label{mainalgo:line:find-featset}}
  \IF {{$(H_j \subseteq H_i)$ $\land$ $(y_j \notin (d_i \cup y_i))$}\label{mainalgo:line:if-featset}}
		\STATE {$featset \gets featset \cup y_j$} \label{mainalgo:line:featset} 
	\ENDIF
\ENDFOR

\STATE $feat,lbl \gets$ $\Sigma_{\downarrow featset}, \Sigma_{\downarrow y_i}$ \label{mainalgo:line:if-featsamples}

\STATE $t \gets$ {\createdt}$(feat,lbl)$

\FOR { n $\in$  $\mathrm{LeafNodes}$(t)\label{algo-decisiontree-line-skf-start}}

	\IF  {{\Label}(n) = 1 \label{algo-decisiontree-tree}  }
	{
		\STATE $\pi \gets$ {\Path}$(t,root,n)$ \COMMENT{A path from root to node $n$ in tree $t$}
		\STATE $f_i \gets f_i \lor \pi$
	}
	\ENDIF
\label{algo-decisiontree-line-skf-end}
\ENDFOR

\FOR {$y_k \in f_i$}
	\STATE $d_k \gets d_k \cup y_i \cup d_i$ \label{algo-decisiontree-line-dependent}
\ENDFOR

\RETURN $f_i,D$

\end{algorithmic}
\end{algorithm}
We now discuss the subroutines {\candidateskf} and {\refine} in detail. 
  
Algorithm~\ref{decision_tree_skf_algo} shows the {\candidateskf} subroutine. {\candidateskf} assumes access to {\createdt} that constructs a decision tree $t$ from labeled data on a set of features $featset$. It uses the ID3 algorithm~\cite{Q86} and we used the Gini Index~\cite{Q86} as the impurity measure.

 In Algorithm~\ref{decision_tree_skf_algo}, line~\ref{mainalgo:line:featset-hi} includes the feature set, \emph{featset}, for $y_i$ in the dependency set $H_i$. Further, line~\ref{mainalgo:line:if-featset} \textit{extends} the features to include all the $y_j$ variables that have the dependency set $H_j$ as a subset of $H_i$, and $y_j$ does not depend on $y_i$ to allow the decision tree to learn over such $y_j$ as well. Line~\ref{mainalgo:line:if-featsamples} selects valuations of feature set and label from samples $\Sigma$, and learns a decision tree. Then, Lines~\ref{algo-decisiontree-line-skf-start}-\ref{algo-decisiontree-line-skf-end} constructs a logical formula as a representation of the decision tree by constructing a disjunction over all paths in the tree that lead to class label \textbf{1}. In line~\ref{algo-decisiontree-line-dependent}, set $d_k$ is updated for variable $y_k$ that appears as a node in decision tree $t$ for $y_i$.

\begin{algorithm}
	\footnotesize
	\caption{{\refine}($\varphi(X,Y),\boldsymbol{f},\sigma,{\totalorder}$)}
	\label{refine_algo}
\begin{algorithmic}[1]
\footnotesize
\STATE $H \gets \varphi(X,Y) \land (X \leftrightarrow \sigma[X])$; $S \gets (Y \leftrightarrow \sigma[Y'])$
\STATE $Ind \gets$ {\maxsatlist}($H,S$) \label{mainalgo:line:maxsat}

\FOR {$y_k \in Ind$}
{
	
	\label{algo:refine:line:repair-start}

	\STATE {$\hat{Y} \gets \emptyset$}\;
	\FOR {{$y_j \in {Y}$}}
	{
		\IF{ {$H_j\subseteq H_k\land \totalorder[index(y_{j})]>\totalorder[index(y_{k})]$} \label{mainalgo:line:if-haty'}}
		{
			\STATE {$ \hat{Y} \gets \hat{Y} \cup y_j$ \label{mainalgo:line:hatY'}}	
		}
		\ENDIF
	}
	\ENDFOR
	
	\STATE {$G_k \gets  (y_k \leftrightarrow \sigma[y'_k]) \land \varphi(X,Y) \land  ({H_k} \leftrightarrow \sigma[H_k]) \land (\hat{Y} \leftrightarrow \sigma[\hat{Y}])$} \label{mainalgo:line:Gk}
	
	\STATE $ret, \rho \gets$ {\checksat}$(G_k)$ \label{algo:refine:line:checksat}

		\IF { $ret = UNSAT$}
		{
 			\STATE $C \gets$ {\findcore}$(G_k)$ \label{algo:refine:line:unsat}
		
			\STATE $\beta \gets \underset{ l \in  C}{\bigwedge} ite((\sigma[l]=1),l,\lnot l)$ \label{refinement formula}

			\STATE $f_k \gets ite((\sigma[y'_k]=1),f_k \land \lnot \beta, f_k \lor \beta) $ \label{refined skolemfunction}

		}
		
		\ELSE
		{

			\FOR { $y_t \in Y \setminus \hat{Y} $\label{algo:refine:line:sat-start}}
			{
				\IF{ $\rho[y_t] \neq \sigma[y'_t] $}
				{
					\STATE $Ind \gets  Ind.Append(y_t)$
				}
				\ENDIF
			}
			\ENDFOR 
			\STATE $\sigma[y_k] \leftarrow \sigma[y'_k]$
		}\label{algo:refine:line:sat-end} 
		\ENDIF 

}
\ENDFOR
\RETURN $\boldsymbol{f}$\ \label{algo:refine:line:repair-end} 

\end{algorithmic}
\end{algorithm}

Algorithm~\ref{refine_algo} represents the {\refine} subroutine. {\refine} assumes access to the following subroutines: 
\begin{enumerate}
	\item {\maxsatlist}: It takes hard constraints and soft constraints as input. It makes a {\maxsat} solver call on a specification containing hard and soft constraints and returns a set of variables corresponding to which the soft constraints are dropped by {\maxsat} solver in order to satisfy the specification.
	\item {\findcore}: It takes a {\unsat} formula as an input and returns unsatisfiable core ({\unsatcore}) of the formula.
\end{enumerate}

 Algorithm~\ref{refine_algo} first attempts to find the potential candidates to repair using {\maxsatlist}. At line~\ref{mainalgo:line:maxsat},  {\maxsatlist} subroutine essentially calls a MaxSAT solver with $\varphi(X,Y) \land (X \leftrightarrow \sigma[X])$ as hard-constraints and $(Y \leftrightarrow \sigma[Y])$ as soft-constraints to find the potential candidates to repair, it returns a list ({\em Ind}) of $Y$ variables such that candidates corresponding to each of the variables appearing in ({\em Ind}) are potential candidates to repair. For each of the $y_k \in$  {\em Ind}, line~\ref{mainalgo:line:if-haty'} computes $\hat{Y}$, which is a set of $y_j$ variable that appears after $y_k$ in {\totalorder} and corresponding $H_j$ is a subset of $H_k$. %

Next, Algorithm~\ref{refine_algo}  checks the satisfiability of the $G_k$ formula at line~\ref{algo:refine:line:checksat}. If $G_k$ is \unsat, line~\ref{algo:refine:line:unsat} attempts to find the {\unsatcore} of $G_k$ using subroutine {\findcore}, and line~\ref{refinement formula} constructs a repair formula $\beta$, using the literals corresponding to unit clauses in {\unsatcore}. Depending on the value of $\sigma[y'_k]$, $\beta$ is used to strengthen or weaken $f_k$ at line~\ref{refined skolemfunction}. If $G_k$ is {\sat} and $\rho \models G_k$, lines~\ref{algo:refine:line:sat-start}-\ref{algo:refine:line:sat-end} look for other potential candidates to repair, and add all $y_t$ variables for which $\rho[y_t]$ is not same as $\sigma[y'_t]$ to the list {\em Ind}.

Note that in line~\ref{mainalgo:line:Gk}, we add a constraint $\hat{Y} \leftrightarrow \sigma[\hat{Y}]$  in $G_i(X,Y)$ where $\hat{Y}$ is a set of $Y$ variables such that for all $y_j$ of $\hat{Y}$, $H_j \subseteq H_i$. Fixing valuations for such $y_j$ variables helps {\dmanthan} to synthesize a better repair for candidate $f_i$. Consider the following example. Let $\forall X \exists^{H_1} \exists^{H_2} \varphi(X,Y)$, where $\varphi(X,Y): (y_1 \leftrightarrow x_1 \oplus y_2)$, $H_1 = \{x_1\}$ and $H_2 = \{x_1\}$. Let us assume that we need to repair the candidate $f_1$, and $G_1(X,Y) = (y_1 \leftrightarrow \sigma[y'_1]) \land \varphi(X,Y) \land (x_1 \leftrightarrow \sigma[x_1])$. As $G_1(X,Y)$ does not include the current value of $y_2$ that led to the counterexample, it misses out on driving $f_1$ in a direction that would ensure $y_1 \leftrightarrow x_1 \oplus y_2$. In fact, in this case repair formula $\beta$ would be empty, thereby failing to repair.

Let us consider an example to discuss {\dmanthan} working in detail.
\begin{figure*}
	\begin{minipage}[b]{0.26\textwidth}
		\centering
		\begin{tabular}{cccccc}\toprule
			$x_1$ & $x_2$ & $x_3$ & $y_1$ & $y_2$ & $y_3$  \\ \midrule
			0 & 0 & 0 & 1 & 1 & 0 \\
			0 & 0 & 1 & 1 & 1 & 1\\
			1 & 1 & 0 & 0 & 0 & 1\\ \bottomrule
			
		\end{tabular}
		\captionof{figure}{\label{tab:samples_F}Samples of $\varphi(X,Y)$}
	\end{minipage}
	\hfill
	\begin{minipage}[b]{0.24\textwidth}

			\centering
			\tikz{
				\node[obs,fill=green!10,minimum size=0.7cm] (x1) {$x_1$};%
				\node[latent, rectangle, minimum width=0.2cm,
				minimum height = 0.5cm, below=of x1, yshift=0.9cm, xshift= 1cm,fill=red!10] (l1) {$0$}; %
				\node[latent, rectangle, minimum width=0.2cm,
				minimum height = 0.5cm, below=of x1, yshift=0.9cm, xshift= -1cm,fill=red!10] (l2) {$1$}; %
				\edge {x1}   {l1}
				\edge {x1}   {l2}
				
				\path[->,draw]
				
				(x1) edge node[above] {$1$} (l1)
				(x1) edge node[above] {$0$} (l2)
				
			}
			\captionof{figure}{\label{decisiontree1F}Decision tree for $y_1$}
		
	\end{minipage}
	\hfill
	\begin{minipage}[b]{0.24\textwidth}

		\centering
		\tikz{
			\node[obs,fill=green!10,minimum size=0.7cm] (x1) {$y_1$};%
			\node[latent, rectangle, minimum width=0.2cm,
			minimum height = 0.5cm, below=of x1, yshift=0.9cm, xshift= 1cm,fill=red!10] (l1) {$1$}; %
			\node[latent, rectangle, minimum width=0.2cm,
			minimum height = 0.5cm, below=of x1, yshift=0.9cm, xshift= -1cm,fill=red!10] (l2) {$0$}; %
			\edge {x1}   {l1}
			\edge {x1}   {l2}
			
			\path[->,draw]
			
			(x1) edge node[above] {$1$} (l1)
			(x1) edge node[above] {$0$} (l2)
			
		}
		\captionof{figure}{\label{decisiontree2F}Decision tree for $y_2$}
		
	\end{minipage}
		\begin{minipage}[b]{\textwidth}

				\centering
				\tikz{
					\node[obs,fill=green!10,minimum size=0.7cm] (x1) {$x_3$};%
					\node[latent,below=of x1, minimum size=0.7cm, yshift=0.9cm, xshift=-1cm,fill=green!10] (x2) {$x_2$}; %
					\node[latent, rectangle,  minimum width=0.2cm,
					minimum height = 0.5cm, below=of x1, yshift=0.9cm, xshift= 1cm,fill=red!10] (l1) {$1$}; %
					\node[latent, rectangle, minimum width=0.2cm,
					minimum height = 0.5cm, below=of x2, yshift=0.9cm, xshift= -1cm,fill=red!10] (l2) {$0$}; %
					\node[latent, rectangle,  minimum width=0.2cm,
					minimum height = 0.5cm, below=of x2, yshift=0.9cm, xshift= 1cm,fill=red!10] (l3) {$1$}; %
					\edge {x1}   {l1}
					\edge {x2}   {l2}
					\edge {x2}   {l3}
					
					\path[->,draw]
					(x1) edge node [above] {$0$} (x2)
					(x1) edge node[above] {$1$} (l1)
					(x2) edge node[above] {$0$} (l2)
					(x2) edge node[above] {$1$} (l3)
					
				}
				
			\captionof{figure}{\label{decisiontree3F}Decision tree for $y_3$}
			
		\end{minipage}
\end{figure*}
We now illustrate Algorithm~\ref{algo:main_algo} through an example.

\begin{example}
	\label{example:main}

	Let $X=\{x_1,x_2,x_3\}$, $Y=\{y_1,y_2,y_3\}$ in $\forall X \exists^{H_1} y_1$ $\exists^{H_2} y_2 \exists^{H_3} y_3 \varphi(X,Y)$ where $\varphi(X,Y)$ is $(x_1 \lor y_1) \land (y_2 \leftrightarrow (y_1 \lor \lnot x_2)) \land (y_3 \leftrightarrow (x_2 \lor x_3))$, and $H_1 = \{x_1\}, H_2 = \{x_1,x_2\}$, and $H_3 = \{x_2,x_3\}$.\\
	
    \textbf{Data Generation}: {\dmanthan} generates training data through adaptive sampling. Let us assume the sampler generates data as shown in Figure \ref{tab:samples_F}.\\ 
    	
    \textbf{Candidate Learning}: As $H_1 \subset H_2$, {\dmanthan} adds a dependency constraint that $y_1$ can not depend on $y_2$. {\dmanthan} now attempts to learn candidates and calls {\candidateskf} for each $y_i$. As $H_3 \not \subseteq H_1$, $y_1$ can not depend on $y_3$, the feature set for $y_1$ only includes $H_1$. The decision tree construction uses the samples of $\{x_1\}$ as features and samples of $\{y_1\}$ as labels. The candidate function $f_1$ is constructed by taking a disjunction over all the paths that end in leaf nodes with label $1$. As shown in Figure \ref{decisiontree1F}, $f_1$ is $\lnot x_1$.
	 
     As $H_1 \subset H_2$, the feature set for $y_2$ includes $H_2$ and $y_1$, however it can not include $y_3$ as $H_3 \not \subseteq H_2$. So, the decision tree construction uses the samples of $\{x_1,x_2,y_1\}$ as features and samples of $\{y_2\}$ as labels. The candidate function $f_1$ is constructed by taking a disjunction over all paths that end in leaf nodes with label $1$: as shown in Figure \ref{decisiontree2F}, $f_2$ is synthesized as $y_1$. Similarity, for $y_3$, the feature set is $H_3$, and a decision tree is constructed as shown in Figure~\ref{decisiontree3F} with samples of $x_2,x_3$ as features and samples of $y_3$ as label. We get $f_3:= x_3 \lor (\lnot x_3 \land x_2)$. At the end of {\candidateskf}, we have $f_1 := \lnot x_1, \; f_2 := y_1, \; f_3 := x_3 \lor (\lnot x_3 \land x_2) \;$. Let us assume the total order returned by {\findorder} is $ {\totalorder} = \{y_3, y_2 ,y_1 \}$.\\
	
    \textbf{Verification}: We construct $E(X,Y') = \lnot \varphi(X,Y') \land (Y' \leftrightarrow \boldsymbol{f})$, which turns out to be {\sat}, and let  $\delta = \langle x_1 \leftrightarrow 1$, $x_2 \leftrightarrow 0$, $x_3 \leftrightarrow 0$,  $y'_1 \leftrightarrow 0$, $y'_2 \leftrightarrow 0$, $y'_3 \leftrightarrow 0 \rangle$ be a satisfying assignment of $E(X,Y')$. Next, {\dmanthan} checks if the $X$ valaution can be extended to $Y$ in order to satisfy the specification. It checks satisfiability of $\varphi(X,Y) \land (x_1 \leftrightarrow \delta[x_1]) \land (x_2 \leftrightarrow \delta[x_2]) \land (x_3 \leftrightarrow \delta[x_3])$, which turns out to be {\sat}, let $\pi$ be the satisfying assignment, $\pi = \langle x_1 \leftrightarrow 1$, $x_2 \leftrightarrow 0$, $x_3 \leftrightarrow 0$, $y_1 \leftrightarrow 1$, $y_2 \leftrightarrow 1$, $y_3 \leftrightarrow 0 \rangle$. Let $\sigma = \langle x_1 \leftrightarrow 1$, $x_2 \leftrightarrow 0$, $x_3 \leftrightarrow 0$, $y_1 \leftrightarrow 1$, $y_2 \leftrightarrow 1$, $y_3 \leftrightarrow 0$, $y'_1 \leftrightarrow 0$, $y'_2 \leftrightarrow 0$, $y'_3 \leftrightarrow 0 \rangle$ be a counterexample to fix.\\
    
     \textbf{Candidate Repair}: In order to find the candidates to repair, {\maxsat} solver is called with  $\varphi(X,Y) \land (x_1 \leftrightarrow \sigma[x_1]) \land (x_2 \leftrightarrow \sigma[x_2]) \land (x_3 \leftrightarrow \sigma[x_3])$  as hard constraints and $(y_1 \leftrightarrow \sigma[y'_1]) \land (y_2 \leftrightarrow \sigma[y'_2]) \land (y_3 \leftrightarrow \sigma[y'_3])$ as soft constraints in {\maxsatlist}. Let {\maxsatlist} returns $ind =\{y_2\}$. 
     
     Repair synthesis commences for $f_2$ with a satisfiability check of $G_2= \varphi(X,Y) \land (x_1 \leftrightarrow \sigma[x_1]) \land (x_2 \leftrightarrow \sigma[x_2]) \land (y_1 \leftrightarrow \sigma[y'_1]) \land (y_2 \leftrightarrow \sigma[y'_2])$. Notice, here we can constrain $G_2$ with $y_1$ as $H_1 \subset H_2$. The formula is unsatisfiable, and {\dmanthan} calls {\findcore}, which returns variable $\lnot x_2$, since the constraints $(x_2 \leftrightarrow \sigma[x_2])$ and  $(y_2 \leftrightarrow \sigma[y'_2])$ are not jointly satisfiable in $G_2$. As the output of candidate $f_2$ for the assignment $\sigma$ must change from 0 to 1, $f_2$ is repaired by disjoining with $\lnot x_2$, and we get $f_2:= y_1 \lor \lnot x_2$ as the new candidate.
     
     The updated candidate vector $\boldsymbol{f}: \langle f_1 := \lnot x_1, \; f_2 := y_1 \lor \lnot x_2, \; f_3 := x_3 \lor (\lnot x_3 \land x_2) \;\rangle$ passes the verification check, that is, the formula $E(X,Y')$ is {\unsat}. Thus, {\dmanthan} returns  $\boldsymbol{f}$ as a Henkin function vector.
\end{example}

By definition of Henkin functions, we know that the following lemma holds: 
\begin{lemma} ${\boldsymbol{f}}$ is a Henkin function vector if and only if 
	$\lnot \varphi(X,Y) \land (Y \leftrightarrow \boldsymbol{f})$ is {\unsat}.%
\end{lemma}

{\dmanthan} returns a function vector only when $E(X,Y'): \lnot \varphi(X,Y') \land (Y'\leftrightarrow \boldsymbol{f})$ is {\unsat}, and each function $f_i$ follows Henkin dependencies by construction. Therefore {\dmanthan} is sound, and returned function vector is a Henkin function vector.

{\bfseries Limitations:} There are instances for which {\dmanthan} might not be able to repair a candidate vector, and consequently is not complete. The limitation is that the formula $G(X,Y)$ (Formula~\ref{eqn:Gformuladqbf}) is not aware of Henkin dependencies. 

Let us consider an example, $\phi: \forall X  \exists^{H_1} y_1$ $\exists^{H_2} y_2  \;\varphi(X,Y)$ where $X = \{x_1,x_2,x_3\},$ $Y = \{y_1,y_2\}$, $\varphi(X,Y) := \lnot (y_1 \oplus y_2)$, $H_1 = \{x_1,x_2\}$, and $H_2 = \{x_2,x_3\}$. Note that $\phi$ is True and Henkin functions are $\boldsymbol{f}:= \langle f_1(x_1,x_2): x_2, f_2(x_2,x_3): x_2 \rangle$. Let us assume the candidates learned by {\dmanthan} is $\boldsymbol{f}:= \langle f_1(x_1,x_2): x_2 ,\; f_2(x_2,x_3):\lnot x_2\rangle$. 
The learned candidates are not Henkin functions as $E(X,Y')$ is {\sat}. Let the counterexample to repair is $\sigma$ is $\langle x_1 \leftrightarrow 0$, $x_2 \leftrightarrow 0$, $x_3 \leftrightarrow 0$, $y_1 \leftrightarrow 0$, $y_2 \leftrightarrow 0$, $y'_1 \leftrightarrow 0$, $y'_2 \leftrightarrow 1 \rangle$. 

Let the candidate to repair is $y_2$, and corresponding $G_2$ formula is $G_2 := \varphi(X,Y) \land (x_2 \leftrightarrow 0) \land (x_3 \leftrightarrow 0) \land (y_2 \leftrightarrow 1)$. As $H_1 \not \subseteq H_2$, the formula $G_2$ is not allowed to constrain on $y_1$. $G_2$ turns out {\sat}, suggesting that we should try to repair $y_1$ instead of $y_2$, but as $y_1$ is also not allowed to depend on $y_2$, the formula $G_1$ would also be {\sat}. Therefore, {\dmanthan} is unable to repair candidate $\boldsymbol{f}$ to fix counterexample the $\sigma$. {\dmanthan} would not be able to synthesize Henkin functions for such a case. Hence, {\dmanthan} is not complete.

\section{Experimental Results }~\label{sec:experimental}
We implemented {\dmanthan}\footnote{{\dmanthan} is available at \url{https://github.com/meelgroup/manthan}} using Python, and it employs Open-WBO~\cite{MML14} for {\maxsat} queries, PicoSAT~\cite{B08} to find {\unsat} cores, ABC~\cite{abc} to represent and manipulate Boolean functions, CMSGen to generate the required samples~\cite{GSCM21}, UNIQUE~\cite{F20} to extract definition for uniquely defined variables, and Scikit-Learn~\cite{sklearn} to learn the decision trees. 

{\bfseries Instances:} We performed an extensive comparison on 563 instances consisting of a union of instances from the DQBF track of QBFEval18, 19, and 20~\cite{qbfeval20}, which encompass equivalence checking problems, controller synthesis, and succinct DQBF representations of propositional satisfiability problems.

{\bfseries Test hardware:} All our experiments were conducted on a high-performance computer cluster with each node consisting of a E5-2690 v3 CPU with 24 cores and 96GB of RAM, with a memory limit set to 4GB per core. All tools were run in a single core with a timeout of 7200 seconds for each benchmark.

{\bfseries Tools compared with:} We performed a comparison vis-a-vis the prior state-of-the-art techniques, {\hqs}~\cite{GWRSSB15} and {\pedant}~\cite{RSS21}. Note that we compared {\dmanthan} with the tools that can synthesize Henkin functions for True DQBF; the rest all the DQBF solvers, including DepQBF~\cite{LB10}, DQBDD~\cite{J20} do not synthesize such functions. The DQBF preprocessor HQSpre~\cite{WRMB17} is invoked implicitly by {\hqs}. We found that the performance of {\pedant} degrades with the preprocessor HQSPre; therefore, we consider the results of {\pedant} without preprocessing. {\dmanthan}  is used without HQSpre.

{\bfseries Evaluation objective:} It is well-known that different techniques are situated differently for different classes of instances in the context of NP-hard problems. The practical adoption often employs a portfolio approach~\cite{DPVV21,HPSS18,XHHL08}. Therefore, in practice, one is generally interested in evaluating the impact of a new technique on the portfolio of existing state-of-the-art tools. Hence, to evaluate the impact of our algorithm on the instances that the current algorithms cannot handle, we focus on the \textbf{Virtual Best Synthesizer} (VBS), which is the portfolio of the best of the currently known algorithms. If at least one tool in the portfolio could synthesize Henkin functions for a given instance, it is considered to be synthesized by VBS; that is, VBS is at least as powerful as each tool in the portfolio. The time taken to synthesize Henkin functions for the given instance by VBS is the minimum of the time taken by any tool to synthesize a function for that instance.

\begin{figure}[h]
	\centering
	\includegraphics[scale=0.40]{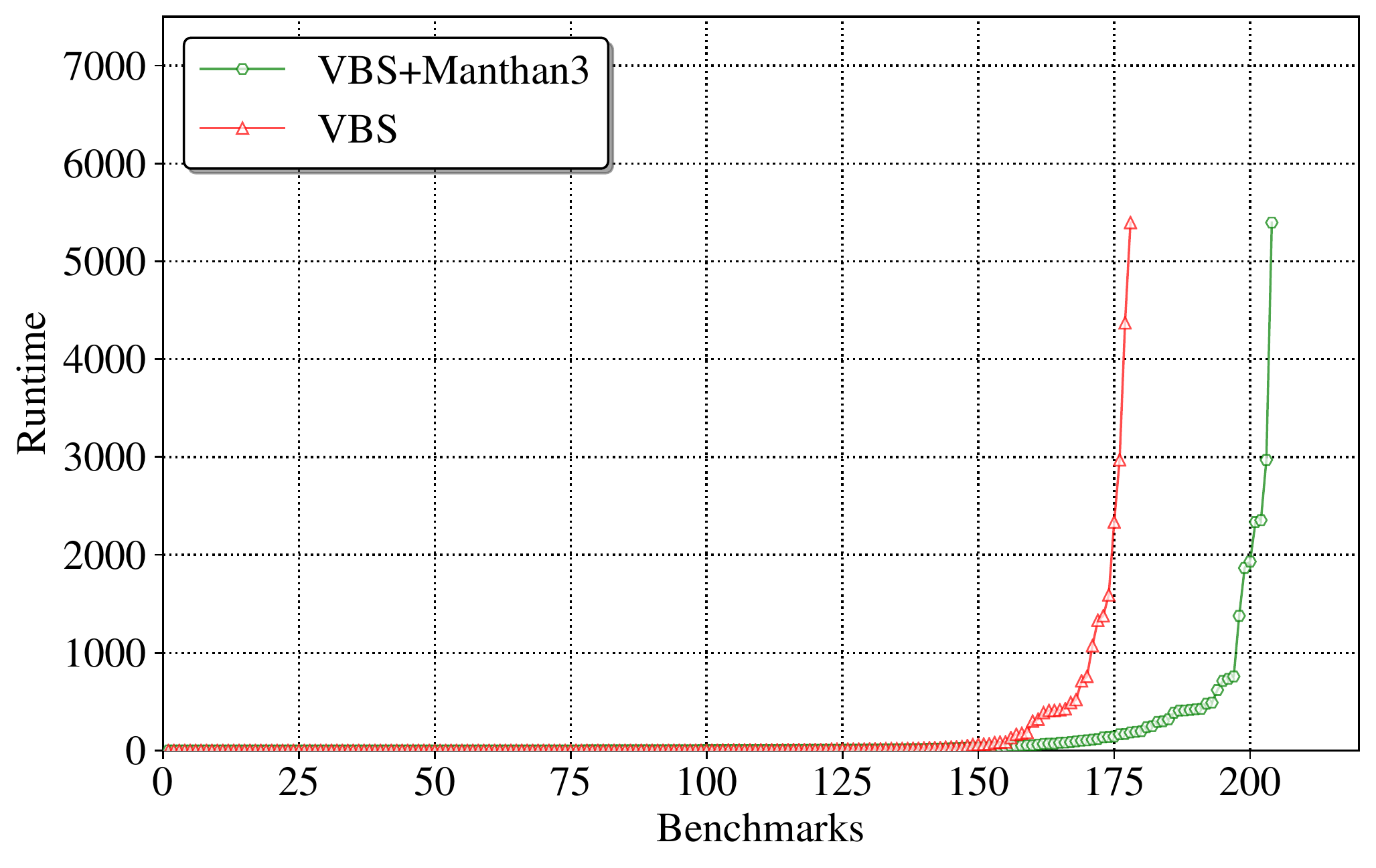}
	\captionof{figure}{ \footnotesize Virtual Best Synthesizing Henkin functions with/without {\dmanthan}. VBS in the plot represents VBS of {\hqs} and {\pedant}. A point  $\langle x,y \rangle$ implies that a tool took less than or equal to $y$ seconds to synthesize a Henkin function vector for $x$ many instances on a total of $563$ instances.}~\label{fig:vbs}
\end{figure}

{\bfseries Results:} Figure~\ref{fig:vbs} represents the cactus plot for VBS of {\hqs} and {\pedant} vis-a-vis with VBS of {\hqs}, {\pedant}, and {\dmanthan}. We observe that the VBS with {\dmanthan} synthesizes functions for 204 instances while VBS without {\dmanthan} synthesizes functions for only 178 instances; that is, the VBS improves by 26 instances with {\dmanthan}. Of 563 instances, for 204 instances, Henkin functions are synthesized by at least one of three tools. {\dmanthan} achieves the smallest synthesizing time on 42 instances, including 26 instances for which none of the other tools could synthesize Henkin functions. 

\begin{figure*}[h]
	\centering
	\begin{minipage}{\textwidth}
		\begin{minipage}[t]{0.45\textwidth}
			\centering
				\includegraphics[scale=0.32]{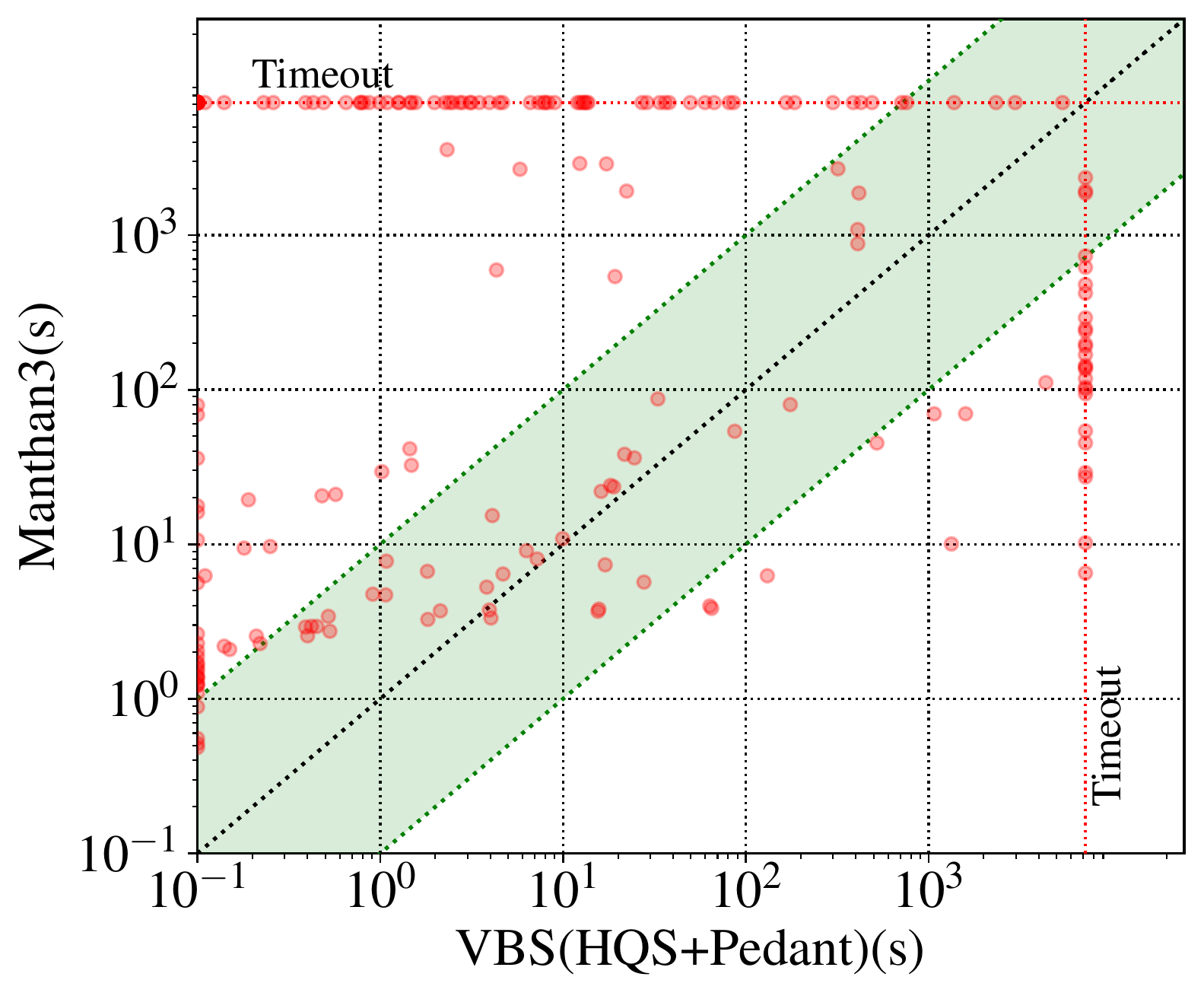}
				\captionof{figure}{\footnotesize {\dmanthan} vs {VBS({\hqs}+{\pedant})}.}\label{fig:d-vs-ph}

		\end{minipage}
		\hfil
		\begin{minipage}[t]{0.45\textwidth}

			\centering
			\includegraphics[scale=0.32]{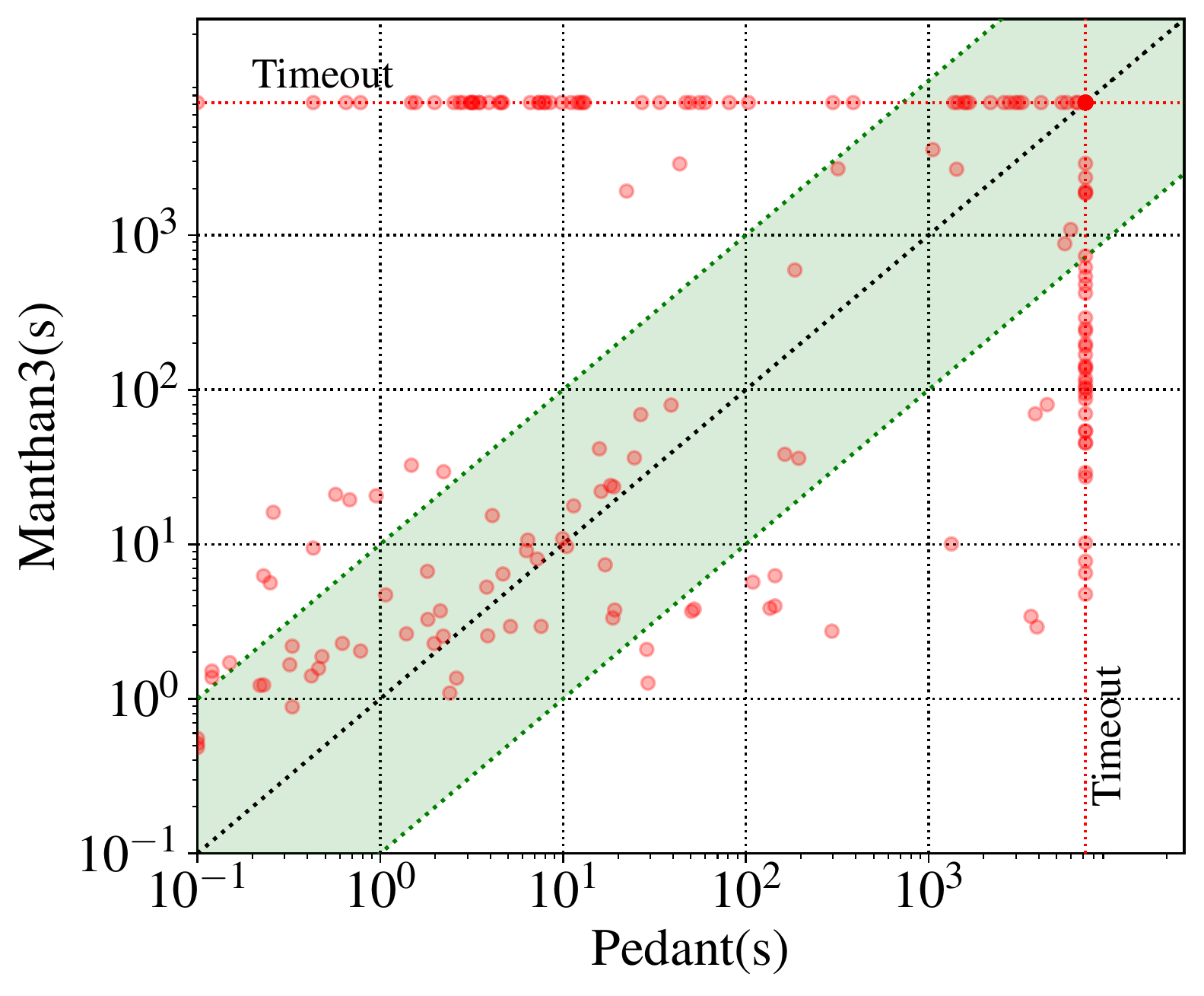}
			\captionof{figure}{\footnotesize {\dmanthan} vs. {\pedant}}\label{fig:pedant-dmanthan}
			
		\end{minipage}
		\hfil
		\begin{minipage}[t]{0.45\textwidth}

						\centering

			\includegraphics[scale=0.32]{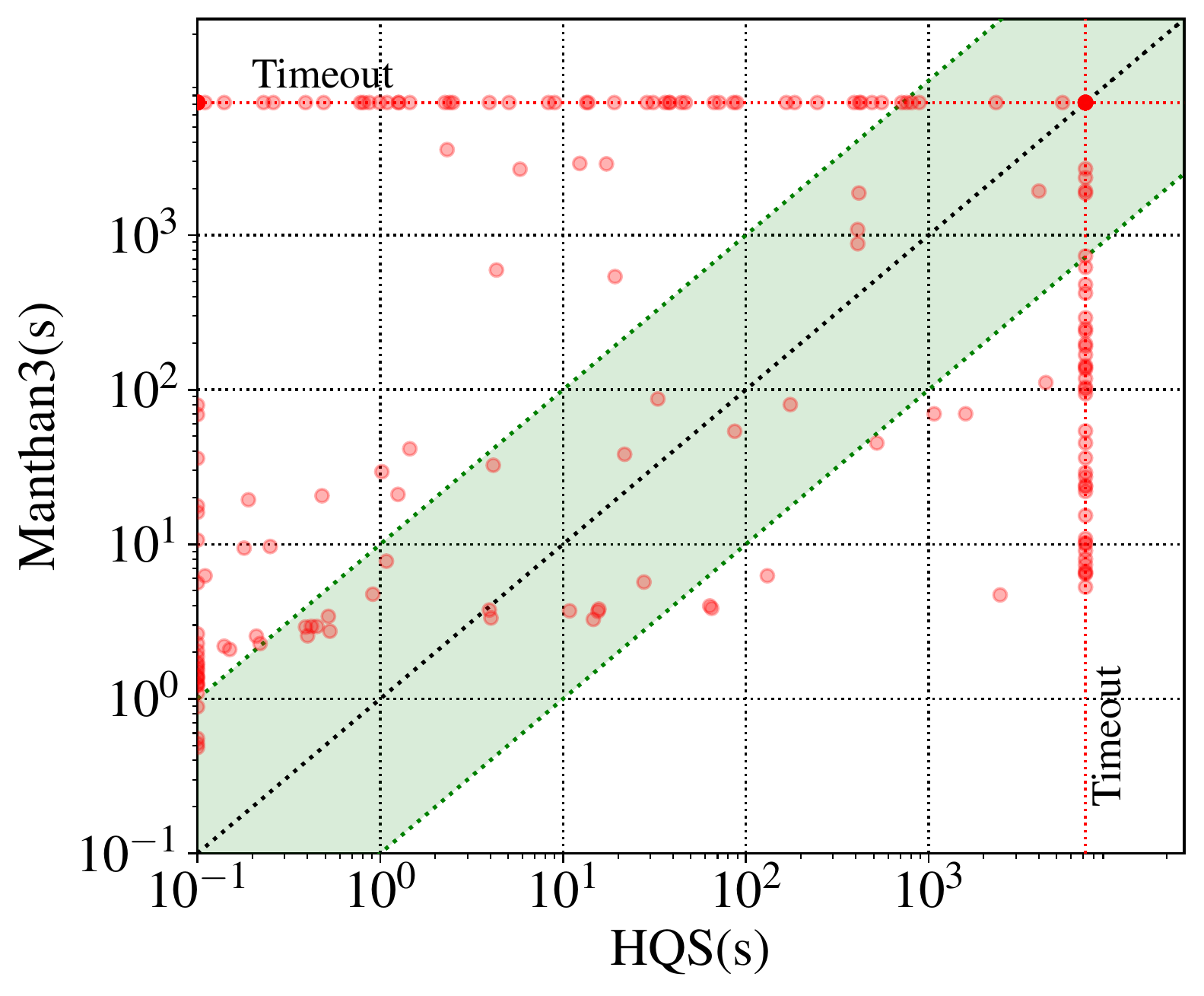}
			
			\captionof{figure}{\footnotesize {\dmanthan} vs. {\hqs}.}\label{fig:hqs-dmanthan}
			
		\end{minipage}
		\hfill
		\begin{minipage}[t]{0.45\textwidth}
			\centering
						\includegraphics[scale=0.32]{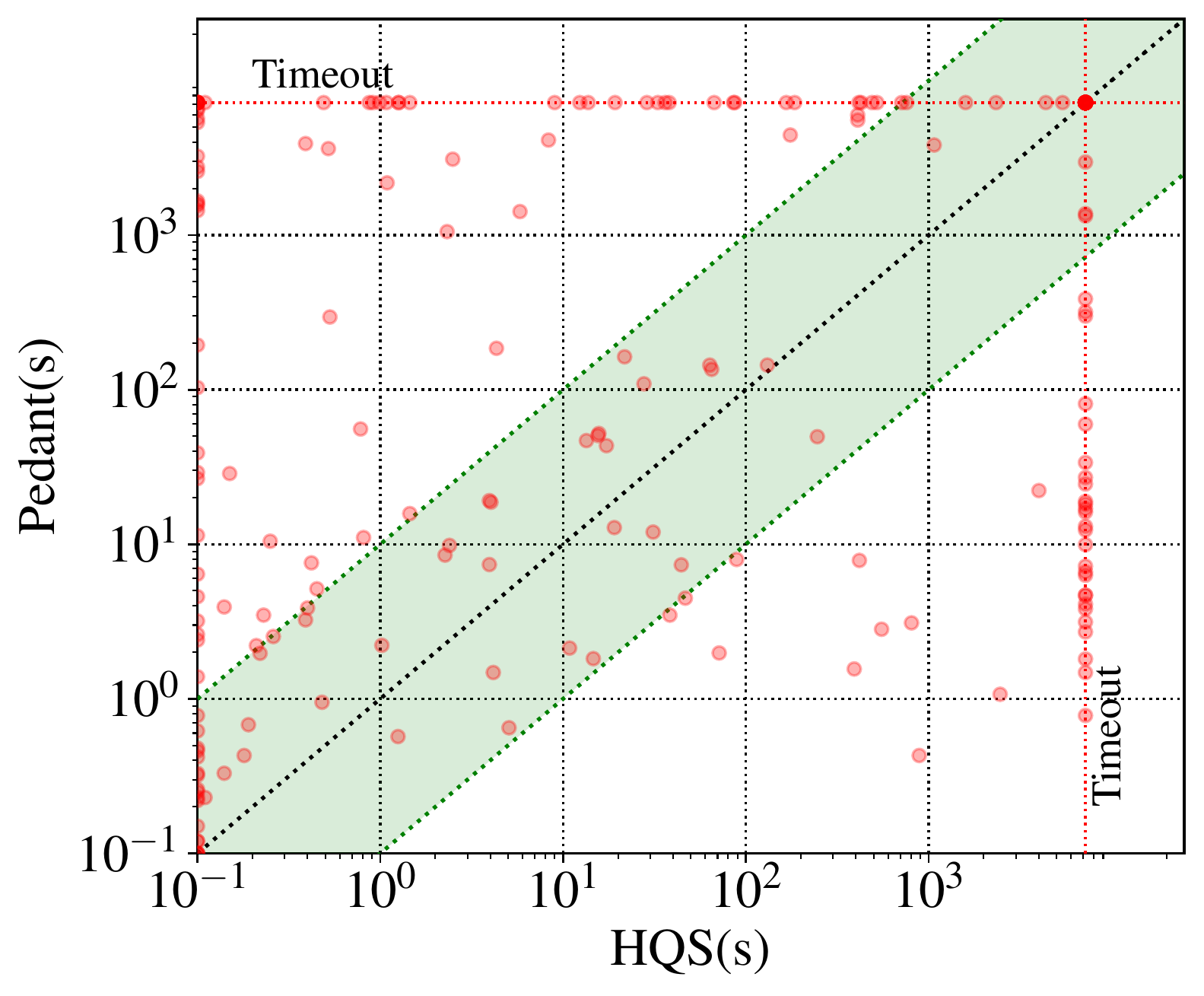}
						\captionof{figure}{\footnotesize  {\pedant} vs. {\hqs}.}\label{fig:ph}
			
		\end{minipage}
	\end{minipage}
	\caption*{\footnotesize A point  $\langle x,y \rangle$ implies that the synthesizer on $\langle x \rangle$ axis took $x$ sec. while the synthesizer on $\langle y \rangle$ axis tooks $y$ sec. to synthesize Henkin functions for an instance.}
\end{figure*}

Figure~\ref{fig:d-vs-ph} highlights that the performance of {\dmanthan} is orthogonal to existing tools. Furthermore, as shown in green area of Figure~\ref{fig:d-vs-ph}, for 47 instances {\dmanthan} took less than or equal to additional 10 seconds to synthesize Henkin functions than by the VBS with {\hqs} and {\pedant}.

Figure~\ref{fig:hqs-dmanthan} (resp. Figure ~\ref{fig:pedant-dmanthan}) represents scatter plot for {\dmanthan} vis-a-vis with {\hqs} (resp. {\pedant}). The distribution of the instances for which functions are synthesized shows that all three tools are incomparable. There are many instances where only one of these tools succeeds, and others fail. 

In total there are 148, 138 and 116 instances for which {\hqs}, {\pedant} and {\dmanthan} could synthesize Henkin functions respectively. Moreover, there are 40 instances for which {\dmanthan} could synthesize Henkin functions, whereas {\hqs} could not. Similarly, there are 37 instances for which {\pedant} could not synthesize Henkin functions and {\dmanthan} synthesized. There are in total $88$ instances for which {\dmanthan} was not able to synthesize functions, however, either {\pedant} or {\hqs} could synthesize Henkin functions.  Due to incompleteness of {\dmanthan}, it could not handle $49$ out of those $88$ instances and for remaining instances it timed out.  

Figure~\ref{fig:ph} shows that there is no best tool even amongst the existing tools, {\pedant} and {\hqs}. Although both tools could synthesize functions for (almost) the same number of instances, the instances belong to different classes. 

The results show that different approaches are suited for different classes of instances, and {\dmanthan} pushes the envelope in Henkin synthesis by handling instances for which none of the state-of-the-art tools could synthesize Henkin functions.

	\section{Conclusion}~\label{sec:conclusion}
Henkin synthesis has wide-ranging applications, including circuit repair, partial equivalence checking, and controller synthesis. In this work, we proposed a Henkin synthesizer, {\dmanthan}, building on advances in machine learning and automated reasoning. {\dmanthan} is orthogonal to existing approaches for Henkin function synthesis that hints that the machine learning-based algorithm employed by {\dmanthan} is fundamentally different from that used by the current Henkin synthesizers. We are interested in understanding these points of deviation better.

\paragraph{Acknowledgements.}  This work was supported in part by National Research Foundation Singapore under its Campus for Research Excellence and Technological Enterprise (CREATE) programme, NRF Fellowship Programme [NRF-NRFFAI1-2019-0004], Ministry of Education Singapore Tier 2 grant [MOE-T2EP20121-0011], and Ministry of Education Singapore Tier 1 Grant [R-252-000-B59-114 ].  The computational work was performed on resources of the National Supercomputing Centre, Singapore \url{https://www.nscc.sg}.

\bibliographystyle{splncs04}
\bibliography{ref}

\begin{thebibliography}{10}
\providecommand{\url}[1]{\texttt{#1}}
\providecommand{\urlprefix}{URL }
\providecommand{\doi}[1]{https://doi.org/#1}

\bibitem{qbfeval20}
{QBF} solver evaluation portal 2020 (2020),
  \url{http://www.qbflib.org/qbfeval20.php}

\bibitem{sklearn}
sklearn.tree.decisiontreeclassifier (2021),
  \url{https://scikit-learn.org/stable/modules/generated/sklearn.tree.DecisionTreeClassifier.html}

\bibitem{AACKRS19}
Akshay, S., Arora, J., Chakraborty, S., Krishna, S., Raghunathan, D., Shah, S.:
  Knowledge compilation for boolean functional synthesis. In: Proc. of FMCAD
  (2019)

\bibitem{ACGKS18}
Akshay, S., Chakraborty, S., Goel, S., Kulal, S., Shah, S.: What’s hard about
  boolean functional synthesis? In: Proc. of CAV (2018)

\bibitem{ACJS17}
Akshay, S., Chakraborty, S., John, A.K., Shah, S.: Towards parallel boolean
  functional synthesis. In: Proc. of TACAS (2017)

\bibitem{BKJ14}
Balabanov, V., Chiang, H.J.K., Jiang, J.H.R.: Henkin quantifiers and boolean
  formulae: A certification perspective of dqbf. Proc. of Theoretical Computer
  Science  (2014)

\bibitem{BJ11}
Balabanov, V., Jiang, J.H.R.: Resolution proofs and skolem functions in {QBF}
  evaluation and applications. In: Proc. of CAV (2011)

\bibitem{B08}
Biere, A.: {PicoSAT} essentials. Proc. of JSAT  (2008)

\bibitem{BKS14}
Bloem, R., K{\"o}nighofer, R., Seidl, M.: Sat-based synthesis methods for
  safety specs. In: Proc. of VMCAI (2014)

\bibitem{CHOP13}
Chatterjee, K., Henzinger, T.A., Otop, J., Pavlogiannis, A.: Distributed
  synthesis for ltl fragments. In: Proc. of FMCAD (2013)

\bibitem{DPVV21}
Dudek, J.M., Phan, V.H.N., Vardi, M.Y.: {ProCount}: Weighted projected model
  counting with graded project-join trees. In: Proc. of SAT (2021)

\bibitem{FTV16}
Fried, D., Tabajara, L.M., Vardi, M.Y.: {BDD}-based boolean functional
  synthesis. In: Proc. of CAV (2016)

\bibitem{FKB12}
Fr{\"o}hlich, A., Kov{\'a}sznai, G., Biere, A.: A dpll algorithm for solving
  dqbf. Proc. POS  (2012)

\bibitem{FKB14}
Fr{\"o}hlich, A., Kov{\'a}sznai, G., Biere, A., Veith, H.: idq:
  Instantiation-based {DQBF} solving. In: Proc. of SAT (2014)

\bibitem{M79}
Garey, M.R.: A guide to the theory of np-completeness. Computers and
  intractability  (1979)

\bibitem{GRSWSB13}
Gitina, K., Reimer, S., Sauer, M., Wimmer, R., Scholl, C., Becker, B.:
  Equivalence checking of partial designs using dependency quantified boolean
  formulae. In: Proc. of ICCD (2013)

\bibitem{GWRSSB15}
Gitina, K., Wimmer, R., Reimer, S., Sauer, M., Scholl, C., Becker, B.: Solving
  dqbf through quantifier elimination. In: Proc. of DATE (2015)

\bibitem{GRM20}
Golia, P., Roy, S., Meel, K.S.: Manthan: A data-driven approach for {B}oolean
  function synthesis. In: Proc. of CAV (2020)

\bibitem{GSRM21}
Golia, P., Slivovsky, F., Roy, S., Meel, K.S.: Engineering an efficient boolean
  functional synthesis engine. In: Proceedings of International Conference On
  Computer Aided Design (ICCAD) (2021)

\bibitem{GSCM21}
Golia, P., Soos, M., Chakraborty, S., Meel, K.S.: Designing samplers is easy:
  The boon of testers. In: Proc. of FMCAD (2021)

\bibitem{GSRM19}
Gupta, R., Sharma, S., Roy, S., Meel, K.S.: {WAPS}: Weighted and projected
  sampling. In: Proc. of TACAS (2019)

\bibitem{H61}
Henkin, L.: Some remarks on infinitely long formulas, infinitistic methods
  (1959)

\bibitem{HSB14}
Heule, M.J., Seidl, M., Biere, A.: Efficient extraction of skolem functions
  from {QRAT} proofs. In: Proc. of FMCAD (2014)

\bibitem{HPSS18}
Hoos, H.H., Peitl, T., Slivovsky, F., Szeider, S.: Portfolio-based algorithm
  selection for circuit qbfs. In: Proc. of CP (2018)

\bibitem{J18a}
Janota, M.: Circuit-based search space pruning in qbf. In: Proc. of SAT.
  Springer (2018)

\bibitem{J18}
Janota, M.: Towards generalization in qbf solving via machine learning. In:
  Proc. of the AAAI (2018)

\bibitem{J09}
Jiang, J.H.R.: Quantifier elimination via functional composition. In: Proc. of
  CAV (2009)

\bibitem{JKL20}
Jiang, J.H.R., Kravets, V.N., Lee, N.Z.: Engineering change order for
  combinational and sequential design rectification. In: Proc. of DATE (2020)

\bibitem{JSCTA15}
John, A.K., Shah, S., Chakraborty, S., Trivedi, A., Akshay, S.: Skolem
  functions for factored formulas. In: Proc. of FMCAD (2015)

\bibitem{KM95}
Krynicki, M., Mostowski, M.: Henkin quantifiers. In: Quantifiers: logics,
  models and computation (1995)

\bibitem{abc}
Logic, B., Group, V.: {ABC}: A system for sequential synthesis and verification
  (2021), \url{http://www.eecs.berkeley.edu/~alanmi/abc/}

\bibitem{LB10}
Lonsing, F., Biere, A.: {DepQBF}: A dependency-aware {QBF} solver. Proc. of
  JSAT  (2010)

\bibitem{LE17}
Lonsing, F., Egly, U.: Depqbf 6.0: A search-based {QBF} solver beyond
  traditional {QCDCL}. In: Proc. of CADE (2017)

\bibitem{MML14}
Martins, R., Manquinho, V., Lynce, I.: {Open-WBO}: A modular {MaxSAT} solver.
  In: Proc. of SAT (2014)

\bibitem{PRA01}
Peterson, G., Reif, J., Azhar, S.: Lower bounds for multiplayer noncooperative
  games of incomplete information. Computers \& Mathematics with Applications
  (2001)

\bibitem{Q86}
Quinlan, J.R.: Induction of decision trees. Proc. of Machine learning  (1986)

\bibitem{R19}
Rabe, M.N.: Incremental determinization for quantifier elimination and
  functional synthesis. In: Proc. of CAV (2019)

\bibitem{RT15}
Rabe, M.N., Tentrup, L.: {CAQE}: A certifying {QBF} solver. In: Proc. of FMCAD
  (2015)

\bibitem{RTRS18}
Rabe, M.N., Tentrup, L., Rasmussen, C., Seshia, S.A.: Understanding and
  extending incremental determinization for {2QBF}. In: Proc. of CAV (2018)

\bibitem{RSS21}
Reichl, F.X., Slivovsky, F., Szeider, S.: Certified {DQBF} solving by
  definition extraction. In: Proc. of SAT (2021)

\bibitem{F20}
Slivovsky, F.: Interpolation-based semantic gate extraction and its
  applications to {QBF} preprocessing. In: Proc. of CAV (2020)

\bibitem{J20}
Síč, J., Strejček, J.: {DQBDD:} an efficient {BDD}-based {DQBF} solver. In:
  Proc. of SAT (2021)

\bibitem{TR19}
Tentrup, L., Rabe, M.N.: Clausal abstraction for {DQBF}. In: Proc. of SAT
  (2019)

\bibitem{WWSB16}
Wimmer, K., Wimmer, R., Scholl, C., Becker, B.: Skolem functions for {QBF}. In:
  Proc. of ATVA (2016)

\bibitem{WRMB17}
Wimmer, R., Reimer, S., Marin, P., Becker, B.: Hqspre--an effective
  preprocessor for {QBF} and {DQBF}. In: Proc. of TACAS (2017)

\bibitem{XHHL08}
Xu, L., Hutter, F., Hoos, H.H., Leyton-Brown, K.: Satzilla: portfolio-based
  algorithm selection for {SAT}. Proc. of JAIR  (2008)

\end{thebibliography}

\end{document}